\documentclass{article}

\usepackage[utf8]{inputenc}
\usepackage[T1]{fontenc}
\usepackage{geometry}
\usepackage[english]{babel}
\usepackage{mathtools}
\usepackage{amsmath}

\newcommand{\oo}{\Omega}

\title{A five dimensional  model with a fifth dimension as fundamental as time in terms of a cosmological approach }
\author{Gizem Şengör, Metin Arık}
\date{Department of Physics, Bogazici University, Bebek, Istanbul, Turkey}
\begin{document}
\maketitle

\hspace{-0.2in}\textbf{Abstract}\\

In five dimensional cosmological models, the convention is to include the fifth dimension in a way similar to the other space dimensions. In this work we attempt to introduce the fifth dimension in a way that a time dimension would be introduced. With such an internal space, we are able to obtain accelerated expansion without introducing dark energy. We obtain a five dimensional flat, meaning both Ricci flat and conformally flat, spacetime into which all relevant four dimensional cosmologies can be locally embedded.  We also argue on the choice of the cosmological frame. The choice that is simplest and most convenient in terms of dimensional analysis, amounts to a linearly expanding universe. 

\section{Introduction}

We considered ourselves to be living in a three dimensional space until Einstein changed our notion of time from a parameter to a dimension to explain electrodynamics of moving bodies and led us to think in terms of a four dimensional spacetime. The number of dimensions has been increasing ever since. With Kaluza$^{[1]}$ and Klein$^{[2]}$ the four dimensions were augmented to five in an attempt to unite electromagnetism and gravity and explain the quantization of electric charge. While we are plainly aware of our four dimensional surroundings, nobody has been able to observe a fifth dimension yet. Obviously extra dimensions are going to be helpful, but one needs to explain their observational absence. In the original Kaluza-Klein theory, the metric coefficients are independent of the fifth dimension, this is also known as the cylinder condition. The internal space is compactified in a natural attempt to explain its lack of observation. 

In time, the spirit of Kaluza-Klein theory grew into questioning of embedding general relativity's solutions into higher dimensions. Higher dimensional theories became more fruitful, by relaxing Kaluza's cylinder condition and allowing components of the metric tensor to depend on the extra dimension. In general it is possible to embed all solutions of general relativity in the general canonical metric, but it is not possible to embed them all in Minkowski metric$^{[3]}$. The standard four dimensional cosmological models are special in that they can be embedded into five dimensional flat spacetimes. It is remarked that this raises the consideration of cosmic dynamics as geometric effects in a five dimensional manifold which is flat or Ricci flat.$^{[4]}$ An example of the later case is Induced Matter Theory which holds the internal dimension to be responsible of ordinary matter, such as dust and radiation. It implies that we can view the universe to be either four dimensional and curved or five dimensional and Ricci flat.$^{[5]}$ The present work on the other hand lies within the former case.

The usefulness of a fifth dimension grew when Randall and Sundrum$^{[6,7]}$ used it to explain the hierarchy problem, which brought forth the concept of brane worlds. Although important steps were made with all these works and many others, it seems that there is still much to be done in order to completely understand internal extra dimensions.  Today the number of dimensions have gone up to eleven or one can also say that they came down from twenty six via superstrings with string or M-theory's quest to understand quantum effects of gravity and unite all fundamental forces.$^{[8]}$

Another thing that is as mysterious as extra dimensions is dark energy. The galaxies were first observed to be receding from each other by Hubble. Today we are certain that our universe is accelerating while expanding, in time$^{[9,10]}$. Yet we do not really know what causes this acceleration. We have come up with the term dark energy as the source, with the name dark to emphasize our uncertainity. The accelerating expansion of the universe enters the cosmological metric as an exponential function of time which is the scale factor of three spacelike dimensions. Perhaps the two concepts, dark energy and extra dimensions, are connected with each other$^{[11]}$. A recent attitude towards an explanation of dark energy is to modify the geometric side of Einstein's equations. One attempt which includes extra dimensions, is brane-world gravity, where at high energies massive modes of graviton dominate, gravity leaks off the brane where its weakening initiates acceleration$^{[12]}$.

In this work we want to approach this jungle of dimensions with purely cosmological concerns. We want to see what happens when we introduce an extra spacelike dimension into the cosmological metric, in the same way that a timelike dimension would be introduced. This way we will be putting forth symmetries between time and the internal space, which brings up the question whether internal space can be as fundamental as time. Our main motivation is curiosity while our second motivation is to see if we can obtain the effects of dark energy from this five dimensional metric without having to introduce the cosmological constant. In the end we will have embedded all relevant four dimensional cosmologies in a flat five dimensional spacetime. This will include the case with an exponential scale factor, attributed to dark energy. However all the different cases with radiation, matter, inflation and dark energy will arise for different values of the parameters. We will also argue on the choice for the cosmological frame. In this respect we will consider transformations involving time and internal space. We will see that the natural choice, in terms of simplicity and dimensional considerations, will give a linearly expanding universe. We will end the discussion by considering a braneworld version of our cosmological frame. 

\section{The Metric and The Einstein Tensor}

The Friedmann-Robertson-Walker metric has the following form

\begin{equation}ds^2=-dt^2+a^2(t)d\Sigma^2\end{equation}
where $d\Sigma^2$ is the metric of three spacelike dimensions all of which have uniform curvature. We use natural units with $c=\hbar=1$. The spacelike sections, being scaled by $a(t)$, expand or contract in time. Therefore the scale factor $a(t)$ is what gives us the dynamics of this four dimensional spacetime. Because all three spatial dimensions have the same scale factor they all change by the same amount, hence this universe expands or contracts isotropically only with time. Here the time is proper time, which is what an observer who sees the universe expand around him measures as time. Since it doesn't have a factor dependent on any of the spacelike dimensions in front of it, it has the same value at every point. In other words the cosmological time is the proper time at every point in this spacetime. The role of time is fundamental here. 

We will consider a metric of the form
\begin{equation}ds^2=f^2(t)g^2(w)[-dt^2+dw^2]+a^2(t)b^2(w)\frac{dx^2+dy^2+dz^2}{(1+\frac{\kappa(x^2+y^2+z^2)}{4})^2}\end{equation}
where $\kappa$ is the curvature of spacelike sections with the values $-1$ for negatively curved, $0$ for flat, $+1$ for positively curved, we can always make a coordinate transformation so that
\begin{subequations}
\begin{align}
dT=f(t)dt\\
dW=g(w)dw\end{align} \end{subequations}
\begin{equation}ds^2=-G^2(W)dT^2+F^2(T)dW^2+A^2(T)B^2(W)\frac{dx^2+dy^2+dz^2}{(1+\frac{\kappa(x^2+y^2+z^2)}{4})^2}.\end{equation}
 Here $T$ may be called the cosmological time because it is the only coordinate that an observer will measure as time. But the value measured will change for different observers at different points in $W$, because we cannot get rid of the factor of $W$ in front of time. We cannot get rid of the factor of time in front of $W$ the internal space either. As such, the role of internal space in this five dimensional universe is as fundamental as the role time plays here. We will carry on our calculations in the coordinates where the metric is as it is in $(2)$.

The observable three spacelike dimensions share the same scale factor and are again isotropic. Here they do not evolve only in time but in $w$ as well. Although our internal space, $w$, is a spacelike dimension, it works as a timelike extra dimension would.

Our basis one forms are
\begin{subequations}
\begin{align}
 e^4&=if(t)g(w)dt, \hspace{0.3cm} i=\sqrt{-1} \\
e^5&=f(t)g(w)dw\\
e^i&=a(t)b(w)\frac{dx^i}{1+\frac{\kappa r^2}{4}} \end{align}
\end{subequations}
and we use the metric $g_{\mu \nu}=diag(1,1,1,1,1)$ with $i=1,2,3$. Using Cartan's formalism we get the curvature two forms to be

\begin{equation}{\oo^i}_j=[\frac{\dot a^2(t)}{a^2(t)f^2(t)g^2(w)}-\frac{b'^2(w)}{b^2(w)f^2(t)g^2(w)}+\frac{\kappa}{a^2(t)b^2(w)}]e^i\wedge e^j \end{equation} 

$${\oo^i}_4=[ \frac{\dot a(t) \dot f(t)}{a(t)f^3(t)g^2(w)}-\frac{\ddot a(t)}{a(t)f^2(t)g^2(w)}+\frac{b'(w) g'(w)}{b(w)f^2(t)g^3(w)}]e^4\wedge e^i $$
\begin{equation} + [\frac{\dot a(t)b'(w)}{ia(t)b(w)f^2(t)g^2(w)}- \frac{\dot a(t)g'(w)}{ia(t)f^2(t)g^3(w)}-\frac{b'(w) \dot f(t)}{ib(w)f^3(t)g^2(w)}]e^5\wedge e^i \end{equation}

$$ {\oo^i}_5=[ \frac{b'(w)\dot a(t)}{ia(t)b(w)f^2(t)g^2(w)}- \frac{b'(w) \dot f(t)}{ib(w)f^3(t)g^2(w)}-\frac{\dot a(t)g'(w)}{ia(t)f^2(t)g^3(w)}]e^4\wedge e^i  $$
\begin{equation}+[\frac{b''(w)}{b(w)f^2(t)g^2(w)}-\frac{b'(w)g'(w)}{b(w)f^2(t)g^3(w)}-\frac{\dot a(t)\dot f(t)}{a(t)f^3(t)g^2(w)}]e^5\wedge e^i \end{equation}

\begin{equation}{\oo^4}_5=[ -\frac{\dot f(t)^2}{f^4(t)g^2(w)}+ \frac{\ddot f(t)}{f^3(t)g^2(w)}+\frac{g'(w)^2}{g^4(w)f^2(t)}-\frac{g''(w)}{g^3(w)f^2(t)}]e^4\wedge e^5 \end{equation}
where differentiation with respect to $w$ and $t$ are denoted as
$$ \dot h=\frac{\partial h}{\partial t} $$
$$  h'=\frac{\partial h}{\partial w} $$
 We get the Riemann tensor $R_{\mu \nu \lambda x}$ from curvature two forms by

$${\oo^\mu}_\nu=\frac{1}{2} {R^\mu}_{\nu \lambda x} e^\lambda \wedge e^x$$
and the components of our Einstein tensor by 
$$G_{\mu \nu}=R_{\mu \nu}-\frac{1}{2}g_{\mu \nu}R$$  where R is the Ricci Scalar $R=g^{\mu \nu}R_{\mu \nu}$. All this gives us the following

$$G_{ii}=-2\frac{b''(w)}{b(w)f^2(t)g^2(w)}+2\frac{\ddot a(t)}{a(t)f^2(t)g^2(w)}+\frac{\dot a^2(t)}{a^2(t)f^2(t)g^2(w)}-\frac{b'^2(w)}{b^2(w)f^2(t)g^2(w)}$$
\begin{equation}+\frac{\kappa}{a^2(t)b^2(w)}-\frac{\dot f^2(t)}{f^4(t)g^2(w)}+\frac{\ddot f(t)}{f^3(t)g^2(w)}+\frac{g'^2(w)}{f^2(t)g^4(w)}-\frac{g''(w)}{f^2(t)g^3(w)}\end{equation}

$$G_{44}=3\frac{\dot a(t) \dot f(t)}{a(t)f^3(t)g^2(w)}+3\frac{b'(w)g'(w)}{b(w)f^2(t)g^3(w)}-3\frac{b''(w)}{b(w)f^2(t)g^2(w)}$$
\begin{equation}3\frac{\dot a^2(t)}{a^2(t)f^2(t)g^2(w)}-3\frac{b'^2(w)}{b^2(w)f^2(t)g^2(w)}+3\frac{\kappa}{a^2(t)b^2(w)}\end{equation}

$$G_{55}=3\frac{\ddot a(t)}{a(t)f^2(t)g^2(w)}+3\frac{\dot a^2(t)}{a^2(t)f^2(t)g^2(w)}+3\frac{\kappa}{a^2(t)b^2(w)}$$
\begin{equation}-3\frac{\dot a(t) \dot f(t)}{a(t)f^3(t)g^2(w)}-3\frac{b'^2(w)}{b^2(w)f^2(t)g^2(w)}-3\frac{b'(w)g'(w)}{b(w)f^2(t)g^3(w)} \end{equation}

\begin{equation}G_{54}=3[\frac{b'(w)\dot a(t)}{ia(t)b(w)f^2(t)g^2(w)}-\frac{b'(w)\dot f(t)}{ib(w)f^3(t)g^2(w)}-\frac{\dot a(t) g'(w)}{ia(t)f^2(t)g^3(w)}  ]\end{equation}

\section{ Vacuum Solutions in 5 Dimensions}

Now let us consider the vacuum solutions for flat spacelike sections, that is solutions to $G_{\mu \nu}=0$ with $\kappa=o$.

From $G_{ii}=0$ we get
\begin{equation}2\frac{\ddot a(t)}{a(t)}+\frac{\dot a^2(t)}{a^2(t)}-\frac{\dot f^2(t)}{f^2(t)}+\frac{\ddot f(t)}{f(t)}=2\frac{b''(w)}{b(w)}+\frac{b'^2(w)}{b^2(w)}-\frac{g'^2(w)}{g^2(w)}+\frac{g''(w)}{g(w)}\end{equation}
The right hand side of this equation is purely $w-$dependent, and the left hand side purely $t-$dependent. The only way these two sides are equal to one another is if they are equal to the same constant $k$. Thus out of $G_{ii}$ we get the following two equations
\begin{equation}2\frac{\ddot a(t)}{a(t)}+\frac{\dot a^2(t)}{a^2(t)}-\frac{\dot f^2(t)}{f^2(t)}+\frac{\ddot f(t)}{f(t)}=k\end{equation}
and
\begin{equation}2\frac{b''(w)}{b(w)}+\frac{b'^2(w)}{b^2(w)}-\frac{g'^2(w)}{g^2(w)}+\frac{g''(w)}{g(w)}=k\end{equation}
With the same reasoning we get from $G_{44}=0$

\begin{equation}\frac{\dot a(t) \dot f(t)}{a(t)f(t)}+\frac{\dot a^2(t)}{a^2(t)}=l\end{equation}
\begin{equation}\frac{b''(w)}{b(w)}+\frac{b'^2(w)}{b^2(w)}-\frac{b'(w)g'(w)}{b(w)g(w)}=l\end{equation}
and from $G_{55}=0$

\begin{equation}\frac{\ddot a(t)}{a(t)}+\frac{\dot a^2(t)}{a^2(t)}-\frac{\dot a(t) \dot f(t)}{a(t)f(t)}=m\end{equation}
\begin{equation}\frac{b'^2(w)}{b^2(w)}+\frac{b'(w)g'(w)}{b(w)g(w)}=m\end{equation}

Thus we have two sets of equations, one set related to $t$ and the other related to $w$. We will solve these two sets first and check whether the solutions satisfy  $G_{54}=0$, which gives
\begin{equation}1-\frac{a(t)\dot f(t)}{\dot a(t) f(t)}=\frac{g'(w)b(w)}{g(w)b'(w)}=constant.\end{equation}

 Let's first look at the set related to $t$, whose solution will give us $a(t)$ and $f(t)$

\begin{equation}2\frac{\ddot a(t)}{a(t)}+\frac{\dot a^2(t)}{a^2(t)}-\frac{\dot f^2(t)}{f^2(t)}+\frac{\ddot f(t)}{f(t)}=k\end{equation}
\begin{equation}\frac{\dot a(t) \dot f(t)}{a(t)f(t)}+\frac{\dot a^2(t)}{a^2(t)}=l\end{equation}
\begin{equation}\frac{\ddot a(t)}{a(t)}+\frac{\dot a^2(t)}{a^2(t)}-\frac{\dot a(t) \dot f(t)}{a(t)f(t)}=m\end{equation}

We can get an equation for $a(t)$ by adding the last two equations,

\begin{equation}\frac{\ddot a}{a}+2\frac{\dot a^2}{a^2}=m+l.\end{equation}

If we consider a solution of the form $a(t)=a_0 e^{\nu t}$ and plug this in $(25)$ we get 

\begin{equation}a(t)=a_0 \exp[{\sqrt{\frac{(m+l)}{3}}t}].\end{equation}

By imposing this solution on equation $(24)$ we obtain

\begin{equation}f(t)=f_0\exp[{\frac{2l-m}{\sqrt{3(m+l)}}t}]\end{equation}

When the solutions $(27)$ and $(26)$ are inserted into equations $(22),(23),(24)$ we find that $(23)$ and $(24)$ are satisfied identically where as $(22)$ imposes the condition
\begin{equation}m+l=k.\end{equation} 

A similar approach to the $w$ related set of equations,

\begin{equation}2\frac{b''(w)}{b(w)}+\frac{b'^2(w)}{b^2(w)}-\frac{g'^2(w)}{g^2(w)}+\frac{g''(w)}{g(w)}=k\end{equation}
\begin{equation}\frac{b''(w)}{b(w)}+\frac{b'^2(w)}{b^2(w)}-\frac{b'(w)g'(w)}{b(w)g(w)}=l\end{equation}
\begin{equation}\frac{b'^2(w)}{b^2(w)}+\frac{b'(w)g'(w)}{b(w)g(w)}=m,\end{equation}
gives
\begin{equation}b(w)=b_0 \exp {[\sqrt{\frac{(m+l)}{3}}w]}\end{equation}
and
\begin{equation}g(w)=g_0\exp {[\frac{\sqrt 3 (2m-l)}{\sqrt {m+l}}w]}\end{equation}
where equation$(29)$ imposes the same condition $m+l=k$. Moreover our solutions imply that
\begin{equation} k=m+l \geq 0\end{equation}
since they each contain a $\sqrt{(m+l)}$ term. With these solutions $G_{54}=0$ is satisfied as well.

Thus the vacuum solutions of our five dimensional metric with flat spacelike sections is

\begin{equation}ds^2=f_0^2g_0^2exp[\frac{4l-2m}{\sqrt{3(m+l)}}t+\frac{4m-2l}{\sqrt{3(m+l)}}w](-dt^2+dw^2)+a_0^2b_0^2exp[2\sqrt{\frac{m+l}{3}}(t+w)][dx^2+dy^2+dz^2]\end{equation}

By redefining parameters
\begin{subequations}
\begin{align}
  M_1=\frac{2l-m}{\sqrt{3(m+l)}}, \\
                      M_2=\frac{2m-l}{\sqrt{3(m+l)}} 
\end{align}
\end{subequations}
and rescaling coordinates we can write our metric in its simplest form as
\begin{equation}ds^2=e^{2(M_1t+M_2w)}[-dt^2+dw^2]+e^{2(M_1+M_2)(t+w)}[dx^2+dy^2+dz^2]\end{equation}

\section{The Effective 4 Dimensional Solutions}
We will now consider the above solution of the vacuum five dimensional spacetime at some $w=w_0$ where $w_0$ is a constant. Such a way of considering four dimensional hypersurfaces along constant internal space amounts to local embedding of four dimensional spacetimes into five dimensions. At $w=w_0$ spacetime metric becomes
\begin{equation}ds^2=f_0^2g_0^2exp[\frac{4m-2l}{\sqrt{3(m+l)}}w_0]exp[\frac{4l-2m}{\sqrt{3(m+l)}}t](-dt^2)+a_0^2b_0^2exp[2\sqrt{\frac{m+l}{3}}w_0]exp[2\sqrt{\frac{m+l}{3}}t][dx^2+dy^2+dz^2]\end{equation}
$f_0g_0exp[\frac{2m-l}{\sqrt{3(m+l)}}w_0]$ is just a constant so we can set it equal to another constant $F_0$. With
$$F_0=f_0g_0exp[\frac{2m-l}{\sqrt{3(m+l)}}w_0],$$
$$A_0=a_0b_0exp[\sqrt{\frac{m+l}{3}}w_0]$$
we can write our solution as
\begin{equation}ds^2=-F_0^2exp[\frac{4l-2m}{\sqrt{3(m+l)}}t]dt^2+A_0^2exp[2\sqrt{\frac{m+l}{3}}t][dx^2+dy^2+dz^2].\end{equation}

To write this in terms of the cosmological proper time consider the following coordinate transformation
\begin{equation}d\tilde t=F_0exp[\frac{2l-m}{\sqrt{3(m+l)}}t]dt.\end{equation}
To simplify the notation we will define 
$$\beta =\sqrt{\frac{m+l}{3}},$$
and
$$\alpha =\frac{3\beta}{2l-m}.$$
With all this our coordinate transformation gives,
\begin{equation}\tilde t=F_0 \alpha e^{[\frac{t}{\alpha}]}\end{equation}
and 
\begin{equation}e^{2\beta t}=(\frac{\tilde t}{F_0 \alpha})^{2\beta \alpha}.\end{equation}
This coordinate transformation has turned our solution into
\begin{equation}ds^2=-d \tilde t^2+A_0^2{\tilde t}^{2\alpha \beta}[dx^2+dy^2+dz^2].\end{equation}
We can always absorb $A_0$ into $\vec r$ by a coordinate transformation. So if we drop the tilde, define $\alpha \beta =n$ our metric in its simplest form becomes
 \begin{equation}ds^2=-dt^2+t^{2n}[dx^2+dy^2+dz^2].\end{equation}

Metric $(44)$ contains all the relevant four dimensional cosmologies with ordinary matter. For $n=\frac{2}{3}$ we have matter dominated universe, for $n=\frac{1}{2}$ we have radiation dominated universe.

Furthermore by setting $m=2l$ in $(38)$ we get
\begin{equation}ds^2=f_0^2g_0^2e^{2\sqrt{l}w_0}[-dt^2]+a_0^2b_0e^{2\sqrt{l}w_0}e^{2\sqrt{l}t}[dx^2+dy^2+dz^2].\end{equation}

Before explaining what we have obtained let us simplify this metric further first. The factor $e^{2\sqrt{l}w_0}$ is just a constant which can be set to $c_0^2$. We can also absorb all the constants into $dt^2$  by the coordinate transformation,
$$d\tau =f_0g_0c_0dt$$
\begin{equation}\frac{\tau-\tau _0}{f_0g_0c_0}=t\end{equation}
and define $a_0b_0c_0exp[-\frac{\tau _0}{f_0g_0c_0}]=A_0^2$ so that we have
\begin{equation}ds^2=-d\tau^2+A_0^2e^{[\frac{2\sqrt{l}}{f_0g_0c_0}\tau]}d\vec{r}^2.\end{equation}
Let us denote $\tau$ by $t$ and set $\alpha=\frac{\sqrt l}{f_0g_0c_0}$, the constant $A_0$ can also be absorbed into $d\vec r$ 
\begin{equation}ds^2=-dt^2+e^{2\alpha t}d\vec{r}^2.\end{equation}
Thus we have obtained an exponential scale factor, a behavior attributed to dark energy with $\alpha=H_0$ where $H_0$ is approximately today's value of Hubble's parameter. 

As such we have shown how it is possible to obtain all relevant four dimensional cosmologies with radiation, matter, inflation and dark energy from our five dimensional metric. Of course each case corresponds to different values of the parameters and we are not yet able to switch from one case to another.

\section{The Curvature and The Weyl Tensors}

\hspace{.6cm}So far we have arrived at a five dimensional spacetime whose four dimensional hypersurfaces correspond to relevant cosmologies. At this point it is important to consider the flatness of the five dimensional model to gain further insight. Therefore we will now calculate the Ricci tensor, which carries information about the ordinary matter contend of the universe, and the Weyl tensor, which informs of the presence of grvitational fields. A zero Ricci tensor corresponds to a Ricci flat metric, and a vanishing Weyl tensor corresponds to a conformally flat metric. A flat metric is the one that is both Ricci flat and conformally flat. 

Components of the Weyl tensor in our convention of  Ricci tensor $R_{\nu \lambda}={R^\mu}_{\nu \lambda \mu}$, metric sign $(-,+,+,+)$, are calculated as

$$C_{\rho \sigma \mu \nu}=R_{\rho \sigma \mu \nu}+\frac{1}{d-2}(g_{\rho \mu}R_{\nu \sigma}-g_{\rho \nu}R_{\mu \sigma}-g_{\sigma \mu}R_{\nu \rho}+g_{\sigma \nu}R_{\mu \rho})$$
\begin{equation}-\frac{1}{(d-1)(d-2)}(g_{\rho \mu}g_{\nu \sigma}+g_{\rho \nu}g_{\mu \sigma})R\end{equation}
where $d$ is the number of dimensions.

For our five dimensional solution, in equation $(35)$,
$$R_{ijij}=[f_0^2g_0^2exp(\frac{4l-2m}{\sqrt{3(m+l)}}t+\frac{4m-2l}{\sqrt{3(m+l)}}w) ]^{-1}(\frac{m+l}{3}-\frac{m+l}{3})=0$$
$$R_{i44i}=[f_0^2g_0^2exp(\frac{4l-2m}{\sqrt{3(m+l)}}t+\frac{4m-2l}{\sqrt{3(m+l)}}w)]^{-1}(\sqrt{\frac{m+l}{3}}\frac{l+m}{\sqrt{3(m+l)}}-\frac{m+l}{3})=0$$
$$R_{i45i}=R_{i54i}=\frac{1}{if^2(t)g^2(w)}[\frac{m+l}{3}-\sqrt{\frac{m+l}{3}}\frac{m+l}{\sqrt{3(m+l)}}]=0$$
$$R_{i55i}=\frac{1}{f^2(t)g^2(w)}[\frac{m+l}{3}-\sqrt{\frac{m+l}{3}}\frac{m+l}{\sqrt{3(m+l)}}]=0$$
$$R_{4545}=\frac{1}{f^2(t)g^2(w)}[-\frac{\dot f^2}{f^2}+\frac{\ddot f}{f}+\frac{g'^2}{g^2}-\frac{g''}{g}]=0$$
all the components of Riemann curvature tensor are zero. Therefore the Ricci Scalar, all components of $R_{\mu \nu}$, and the Weyl tensor for the Ricci flat five dimensional metric are all zero. Our five dimensional universe is Ricci flat, meaning it contains no energy nor momentum density, and conformally flat, it doesn't contain any gravitational fields either, in short it is flat and empty.

The Ricci flatness of the metric does not guarantee that it will be conformally flat. It is possible to have Ricci flat solutions with nonzero $R_{\rho \sigma \mu \nu}$. Our universe turned out to be conformally flat because all of its $R_{\rho \sigma \mu \nu}$ vanish. 

It is a well established fact that the Friedmann-Robertson-Walker (FRW) metric can be put in a conformally flat form$^{[13,14]}$. It has been further pointed out that$^{[15,16]}$  calculations on the age of the universe and its matter density carried out in conformally flat spacetime (CFS) coordinates agree better with the observations then those carried out in FRW coordinates. With such emphasis on the conformal flatness of our universe, it is an achievement to be able to embed standard four dimensional conformally flat cosmology in a five dimensional flat spacetime in this work on higher dimensional cosmologies.

\section{Transformations involving internal space and time}
\hspace{.6cm}We now wish to consider SO(1,1) transformations of the $t$ and $w$ coordinates in the form of metric $(37)$ which leave $[-dt^2+dw^2]$ interval invariant. This is the usual Lorentz transformation with a parameter $\alpha$, a boost along $w$ where $t$ and $w$ are transformed as
\begin{subequations}
\begin{align} \tilde t=(cosh\alpha)t+(sinh\alpha)w\\
                      \tilde w=(sinh\alpha)t+(cosh\alpha)w. \end{align}
\end{subequations}
Of course we would like to express the general parameter $\alpha$ in terms of the parameters of our metric. The hyperbolic functions are obliged to satisfy the following identity
\begin{equation}(cosh\alpha)^2-(sinh\alpha)^2=1.\end{equation}
If we define $cosh\alpha$ and $sinh\alpha$ as
\begin{subequations}
\begin{align}cosh\alpha=\frac{M_1}{\sqrt{(M_1^2-M_2^2)}}\\
sinh\alpha=\frac{M_2}{\sqrt{(M_1^2-M_2^2)}}\end{align}
\end{subequations}
the identity is satisfied. Therefore the rapidity for our spacetime, defined in terms of the parameters that appear in our metric is
$\alpha=cosh^{-1}[\frac{M_1}{\sqrt{(M_1^2-M_2^2)}}]$. As such metric $(37)$ transforms into
\begin{equation}ds^2=e^{2\sqrt{(M_1^2-M_2^2)}\tilde t}[-d\tilde t^2+d\tilde w^2]+e^{2\sqrt{(M_1^2-M_2^2)}(\tilde t+\tilde w)}[dx^2+dy^2+dz^2].\end{equation}

Apparently we can remove the $w-$dependent part of the scale factor in front of $[-dt^2+dw^2]$ by a boost along $w$. The scale factor of three spatial dimensions which depends on $(t+w)$ continues to do so in the form of $(\tilde t+\tilde w)$ with only a change in the coefficient, hence $(t+w)$ is a lightlike coordinate. 

We can redefine $M_1^2-M_2^2$ as $\mu^2$ and write this metric as
\begin{equation}ds^2=e^{2\mu\tilde t}[-d\tilde t^2+d\tilde w^2]+e^{2\mu(\tilde t+\tilde w)}[dx^2+dy^2+dz^2].\end{equation}
 This is as if we have set $M_2$=0 via a transformation. At this point we would like to point out that setting the parameters $M_1$ and $M_2$ to certain values amounts to choosing different frames. Of course these frames aren't all equivalent because we will pick out one of them to be the cosmological frame, whose time dimension will be the time referred to as cosmological time. The naive choice is the one in which the scale factor of time is unity. This frame is among those where $M_2=0$ because the scale factor of time here, as in metric $(54)$ can be set to one by the following coordinate transformation
\begin{equation}d\tau=e^{\mu \tilde t}d\tilde t\end{equation}
which makes $e^{\mu \tilde t}=\mu \tau$. We will drop the tilde on $w$ from now on and write the metric in these coordinates 
\begin{equation}ds^2=-d\tau^2+\mu^2 \tau^2dw^2+\mu^2\tau^2e^{2\mu w}[dx^2+dy^2+dz^2].\end{equation}
As such the dimensions of $[\tau]$, $[w]$ and the three space coordinates $[x]$,$[y]$,$[z]$ are all equal to length where as, $[\mu w]$, being the variable of the exponential function, is dimensionless. This form of the metric is appropriate as far as the dimensions are concerned. $\tau$ is the cosmological time and we pick this frame as the cosmological frame. In time this universe expands linearly and it does not contain dark energy. So in a sense our choice of the cosmological frame, is the simplest cosmological case. On the other hand we were able to obtain dark energy for the four dimensional slice from metric $(38)$ by setting $m=2l$, which corresponds to $M_1=0$, in section 4. As we have argued, metric $(37)$ and $(56)$ are different frames, the time coordinate in one is not the same as the time coordinate of the other. If there is a physical reason for us to choose the time of $(37)$ as the cosmological time, it may be the dark energy.

It may be particularly interesting to discuss a braneworld model version of metric $(56)$. The metric can be written as
\begin{equation}ds^2=-d\tau^2+\mu^2 \tau^2dw^2+\mu^2\tau^2e^{-2\mathcal{M}|w|}[dx^2+dy^2+dz^2].\end{equation}
Here we have introduced a brane at $w=0$ and chose $\mu$ to be negative, $\mu=-\mathcal M$. A straigh forward calculation gives us the matter content of this spacetime, by setting $G_{\mu \nu}=\kappa T_{\mu \nu}$, to be
\begin{subequations}  
\begin{align}
p=-4\frac{\delta(w)}{\kappa \mathcal M \tau^2}\\
\rho=6\frac{\delta(w)}{\kappa \mathcal M \tau^2}
\end{align} \end{subequations}
where  $\kappa$ is the five dimensional gravitational constant. We like to point out that as such our world is confined to the brane at $w=0$ and the bulk is empty. The equation of state $\frac{p}{\rho}=-\frac{2}{3}$ satisfies the equation of state for a cosmic wall, which is the brane. Moreover due to our choice of negative $\mu$, as $w$ increases our world at $w_0$ brane becomes the largest in size where as all other worlds are smaller by a factor of $e^{-\mathcal M |w|}$. 

\section{Conclusion}
\hspace{.6cm}As we pointed out in the beginning, it has been shown that  four dimensional curved spacetimes can be embedded in five dimensional flat or Ricci flat spacetimes$^{[4]}$. The former branch has been well studied in literature. It is stated that a matter and radiation dominated four dimensional universe can be embedded in a five dimensional vacuum universe$^{[17,18]}$ and the accelerated expansion of the universe can be obtained via extra dimensional models$^{[19]}$. In this work we have obtained all relevant cosmologies, including dark energy dominated cosmology, as four dimensional slices of a flat, five dimensional metric. We were able to do this by allowing the internal dimension to be fundamental, like time. We name the internal space as fundamental because it affects all the scale factors including that of time. Moreover, although the internal space is a spacelike dimension, the linear combinations of time and the internal space may transform as lightlike coordinates. 

We should like to point out that this paper does not present a detailed cosmological model. In standard four dimensional cosmology, the equation of state that governs the expansion of the universe with time, changes for physical reasons. The early universe goes through different phases, starting with radiation dominated moving on to dust dominated, and so forth by a power law, $a(t)=t^n$, where the value of $n$ changes from one era to another with time. In our model the change of $n$ may be obtained by a pseudo rotation involving internal space and time. Although we can embed all these cases into the same five dimensional metric we are not able to switch from one to another because our parameter $n$ does not depend on time. To turn our model into a physically better suited one it would be necessary to find a different reason for the changes of the time variable to explain this change of $n$ with time which gives the correspondence with radiation dominated, dust dominated, dark energy dominated eras.

We have pointed out that fixing the free parameters amounts to choosing different frames. We have picked out the cosmological frame to be the one in which the scale factor of time is unity and this frame gave us a linearly expanding universe. Thus from dimensional arguments we have shown that,  in a universe where the only dimensional constant is the speed of light, the preferred change of scale size in time amounts to linear expansion. This inevitably brings to mind the present relationship between Hubble's parameter $H_0$, and life time of the universe $t_0$, being $H_0t_0=1$. If this relationship is valid for all times that would indicate a linearly expanding universe. With this in mind our choice of the cosmological frame might indeed be the suitable choice. On the other hand, among the possible frames, the ones that contain dark energy are more complex and have $w$ dependent scale factors for time. A physical reason to choose one of these as the cosmological frame would be dark energy.

We have also discussed a braneworld version of our cosmological frame by putting a brane at $w=0$. Our world turns out to be a linearly expanding universe, confined to the brane. It is also the largest universe in size. The other worlds at different $w-$branes are also linearly expanding universes, and the Hubble time is the same for all of them. The only difference between our universe and these others is that their scale size is smaller by a factor of $e^{-\mathcal M |w|}$.  Dvali $et. al.^{[20]}$ presents a mechanism by which the correct four dimensional gravitational potential may be obtained for static 3-branes embedded in five dimensional Minkowski space.  One of the possible cases to where their mechanism can be applied to consists of matter fields confined to the brane. In our braneworld scenario the 3-brane is dynamic yet the matter fields are still confiend to the brane. Therefore it may be possible to apply the same mechanism here and obtain an expression for the four dimensional gravitational potential with cosmic dynamics.

A Ricci flat spacetime is empty in terms of matter, meaning it contains no pressure or energy density. A vanishing Weyl tensor, which represents the conformal flatness, points the absence of gravitational fields. Therefore a flat universe must be both Ricci flat and Conformally flat, containing neither matter nor gravitational fields.  The ability to embed conventional four dimensional cosmologies in Ricci flat five dimensional spacetimes not only presents a simpler frame to consider the situation in but also gives a geometric explanation of observed material effects, this has been the subject of the briefly mentioned literature. Conformal flatness, on the other hand, is one of the key properties of standart cosmology. Our five dimensional spacetime in which all relevant cosmologies can be locally embedded, is both Ricci flat and conformally flat. Therefore we have achieved the embedding of all relevant four dimensional cosmologies in a flat five dimensional spacetime. 

 The common intuition would be to imagine a four dimensional space expanding along time. Instead what we have introduced here is the three dimensional space expanding along both time and internal space. So in a sense we should visualize this as a four dimensional spacetime evolving along the extra dimension.

We would like to thank Dr. Nihan Katırcı for help with checking some of the calculations on Maple. This work was supported in part by the Turkish Academy of Sciences and Bogazici University project BAP $7128$.

\section{References}

\hspace{0.5cm}[1] T. Kaluza, \textsl{"On the Problem of Unity in Physics"}, 8th International Schol of osmology and Gravitation "Ettore Majorana", Erice, Italy, pp.427-433 (1982)

[2] O. Klein, \textsl{" Quantum Theory and Five-Dimensional Theory of Relativity"}, Z.Phys. 37 (1926) 895-906

[3] P. S. Wesson, \textsl{"The Embedding of General Relativity in Five Dimensional Canonical Space: A Short History and a Review of Recent Physical Progress"},  arXiv:1011.0214 [gr-qc], 2010

[4] M. Lachieze-Rey, \textsl{"The Friedmann-Lemaître models in perspective: Embeddings of the Friedmann-Lemaitre models in flat 5-dimensional space}, Astron. Astrophys. 364, 894-900 (2000)

[5] P. S. Wesson, \textsl{"Space-Time-Matter Modern Higher-Dimensional Cosmology}, World Scientific Publishing 2007

[6] L. Randall and R. Sundrum, \textsl{"A large Mass Hierarchy from a Small Extra Dimension"}, Phys.Rev.Lett.83:3370-3373,1999, hep-th/9905221

[7] L. Randall and R. Sundrum, \textsl{"An Alternative to Compactification"},  Phys.Rev.Lett.83:4690-4693,1999, hep-th/9906064

[8] P. Horava and E. Witten \textsl{" Heterotic and Type I string Dynamics From Eleven Dimensions"}, Nucl.Phys.B460:506-524,1996

[9] Adam G. Riess $et. al.$, \textsl{"Type Ia Supernova Discoveries at z>1 From the Hubble Space Telescope: Evidence for Past Deceleration and Constraints on Dark Energy Evolution"}, Astrophys.J.607:665-687,2004

[10] R. A. Knop $et. al.$, \textsl{"New Constraints on $\Omega_M$, $\Omega_\Lambda$, and w from an Independent Set of Eleven High-Redshift Supernovae Observed with HST"}, Astrophys.J.598:102 (2003)

[11] Je-An Gu, \textsl{"A way to the dark side of the universe through extra dimensions"}, astro-ph/0209223

[12] R. Maartens, \textsl{"Dark Energy from Brane-world Gravity"}, arXiv:astro-ph/0602415

[13] L. Infeld and A. Schild, \textsl{"A New Approach to Kinematic Cosmology"}, Phys.Rev. 68, 250-272 (1945)

[14] O. Gron and S. Johannesen, \textsl{"FRW Universe Models in Conformally Flat Spacetime Coordinates I: General Formalism"}, Eur.Phys.J.Plus 126 (2011) 28

[15] G. Endean, "Redshift and the Hubble Constant in Conformally Flat Spacetime", The Astrophysical Journal 434, 397-401 (1994)

[16] G. Endean, " Cosmology in Conformally Flat Spacetime", The Astrophysical Journal 479, 40-45 (1997)

[17] J. Ponce de Leon, \textsl{"Modern cosmologies from empty Kaluza-Klein solutions in 5D"}, JHEP03(2009)052

[18] P. S. Wesson, \textsl{"An embedding for the big bang"}, Astrophysical Journal, Part 1 436, 547-550 (1994)

[19] E. A. Leon $et.  al.$, \textsl{"Higher Dimensional Cosmology: Relations Among the Radii of Two Homogeneous Spaces"}, Mod. Phys. Lett. A 26, 805 (2011)

[20] G. Dvali, G. Gabadadze, M. Porrati, \textsl{"4D gravity on a brane in 5D Minkowski space"}, Physics Letters B 485 (2000) 208-214

\end{document}